# Influence of Mn doping on the microstructure and

# optical property of ZnO

K.Rajendran[a], S.Banerjee[b], S.Senthilkumaar[c]*, T.K.Chini[b], V.Sengodan[d],

[a] Department of Electronics and Communication Systems, Sri Krishna Arts and Science College, Coimbatore, India

[b] Surface Physics Division, Saha Institute of Nuclear Physics, 1/AF Bidhannagar, Kolkata, India

[c] Department of Chemistry, Faculty of Engineering, PSG College of Technology, Coimbatore, India

[d] Department of Electronics, SNR Sons College, Coimbatore, India

## Abstract

Undoped and Mn doped ZnO samples with different percentage of Mn content (1 mol%, 2 mol% and 3mol%) were synthesized by a simple solvo-thermal method. We have studied the structural, chemical and optical properties of the samples by using x-ray diffraction (XRD), scanning electron microscopy (SEM), energy dispersive x-ray (EDX) analysis, Fourier transform infrared (FTIR) spectroscopy and UV-VIS spectroscopy. The XRD spectra show that all the samples are hexagonal wurtzite structures. The lattice parameters calculated for the Mn doped ZnO from the XRD pattern were found to be slightly larger than those of the undoped ZnO, which indicates substitution of Mn in ZnO lattice. SEM photograph shows the grain size of undoped ZnO is bigger than the Mn doped ZnO's indicating hindrance of grain growth upon Mn doping. As the Mn doping increases the optical band gap decreases for the range of Mn doping reported here.

*Corresponding author E-mail:sskumaarpsg@gmail.com

# 1. Introduction

ZnO materials have received broad attention due to their well-known performance in electronics, optics and photonics [1, 2]. The device application of micro and nanostructure of ZnO is one of the major focuses among researchers to diminish the size of the device in order to achieve higher speeds in its electrical transport and also to study the effect of confinement on optical properties. ZnO is a direct wide band gap ($E_g \sim 3.3$ eV at room temperature) n-type semiconductor with optical transparency in the visible range. This provides opportunities to develop transparent electronics, optoelectronics and integrated sensors [3].

The interest in doping ZnO is to explore the possibility of tailoring its electrical, magnetic and optical properties [4-8]. Such films could be used in areas like electronics, optoelectronics and could be a potential challenger to other oxides such as $SnO_2$ and Indium-Tin Oxides (ITO). The transition metal doped ZnO has the potential to be a multifunctional material with coexisting magnetic, semi-conducting and optical properties [9]. The samples can be synthesized in the bulk and thin film forms and a wide range of magnetic properties including room temperature ferromagnetism have been reported [10-11] and can be applied to short-wave magneto-optical devices [12 ].

The doping of transition metal elements into ZnO offers a feasible means of fine tuning the band gap to make use as UV detector and light emitters. It has been observed earlier by other groups [13,14] that upon Mn doping in ZnO the band gap reduces for low concentration doping (< 3 mol% of Mn) and for higher concentration ( > 3 mol%) the band gap increases as expected on the basis of virtual crystal approximation (VCA) because of the band gap of the MnO $\sim$ 4.2 eV. In many dilute magnetic semiconductor (DMS) systems such deviation from the linear monotonic increase in the form of "band gap bowing" has been observed [15-21]. For the low concentration Mn doping the reduction in the band gap has been theoretically explained as a consequence of exchange interaction between d electrons of the transition metal ions (Mn) and the s and p electron of the host band.

The aim of this work is to evaluate the effect of Mn doping on the microstructure, grain growth and the optical properties of the powder grown by simple solvo-thermal method and thin films prepared by dip coating method for the range of Mn doping ≤ 3 mol%. The EDX analysis shows the excellent oxide formation, wherein dopant ions are present in the host crystal lattice. The FTIR analysis shows the stretching vibrations and the oxide formations. We have measured the energy gap of the different concentration of Mn doped ZnO samples from the UV-VIS absorption spectra and it shows the decrease in band gap upon Mn doping as expected.

## 2. Experimental details:

### 2.1    Materials

The starting materials of analytical grade were zinc acetate dihydrate and manganese acetate tetrahydrate supplied by Lancaster and used as such. Reagent grade absolute ethanol was obtained from Changshu Yangyuan chemical, China and triply distilled.

### 2.2    Synthesis

The experimental flow chart for the synthesis of ZnO and Mn doped ZnO is shown in Fig.1. Undoped ZnO and different concentration of Mn doped ZnO were synthesized through a simple solvo-thermal process (this is also referred as sol-gel by some authors). The undoped ZnO was prepared using zinc acetate $Zn(CH3COO)_2 \cdot 2H_2O$ (14g) and ethanol (sample 'a'). For preparing Mn doped ZnO the zinc acetate $Zn(CH3COO)_2 \cdot 2H_2O$ (14g) and 1mol%, 2mol%, 3mol.% of manganese acetate $Mn(CH3COO)_2 \cdot 4H_2O$ were taken in a round bottom flask and ethanol was slowly added with constant stirring and labeled as sample 'b', 'c' and 'd' respectively. All the mixers were stirred vigorously for 12h and the solution obtained was ultrasonicated for 30 min. Thin films were prepared on the cleaned quartz substrate by dipping in the sol with constant retrieving rate of 0.16mm/sec and were annealed at $500^{o}C$ for 1h.

## 2.3    Characterization

X-ray analyses of the sintered samples were carried out using $CuK_{\alpha}$ radiation on a powder diffractometer (Schimadzu, Model XRD6000) in the $2\theta$ range between $10^{o}$ and $80^{o}$. The chemical compositions and the microstructure of the samples were carried out using energy dispersive x-ray (EDX) analyzer attached with a scanning electron microscope (SEM), model FEI QUANTA 200F. Fourier transform infrared spectroscopy (FTIR) absorption was measured for KBr supported samples over the frequency range of 4000-400 $cm^{-1}$ and at a resolution of 4 $cm^{-1}$, using a model SHIMADZU, FTIR-8400S. Optical transmittance spectra at room temperature of the films have been recorded in the wavelength range of 200 to 2500 nm using a double beam UV-VIS-NIR Spectrophotometer JASCO, model V-570.

## 3. Result and discussion

Powder x-ray diffraction patterns of undoped ZnO and Mn doped ZnO are shown in Fig.2. The characteristic peaks with high intensities corresponding to the planes (100), (002), (101) and lower intensities at (102), (110), (103), (200), (112) and (201) indicate the annealed product is of high-purity hexagonal ZnO wurtzite structure. It's evident from the XRD data that there are no extra peaks due to manganese metal, other oxides or any zinc manganese phase, indicating that the as-synthesized samples are single phase. The Mn ion was understood to have substituted the Zn site without changing the wurtzite structure at $500^{o}C$.

The peaks of the diffraction patterns of doped samples are slightly shifted to left as compared to the undoped ZnO. This shows that small variation in the lattice parameters occur as Mn concentration in the sample increases. Table 1 shows that the lattice constants of Mn doped ZnO were slightly larger than those of undoped ZnO, because the ionic radius of Mn(II) (0.66) is larger than that of Zn(II) (0.60) [12]. The length of both $a$ and $c$ axis expand monotonously with the increasing Mn doping in ZnO. The expansion of the lattice constants and the slight shift of XRD peaks of different concentration of Mn doped ZnO indicated that manganese has really doped into the ZnO structure. The average crystal size (D) was estimated using the Scherrer formula. The crystal size of the undoped ZnO decreases on doping 1 mol% of Mn and on subsequent doping shows an increasing tendency as shown in table 2.

Energy dispersive x-ray analysis showed that the amount of Mn element in the sample increased depending on the increasing Mn incorporation in the solution. As a result, Mn incorporation has a strong effect on the optical, structural and morphological properties of ZnO. The Energy dispersive x-ray analysis shown in Fig.3 (a) consists of Zn and O and fig (b)-(d) confirms the presence of manganese in the ZnO particles and wt% are very nearly equal to the nominal value of Mn in ZnO (Note: traces of Al and C less than 0.1 wt% are observed. The Al peak may be arising from the sample stub). The table 3 list out the wt% of the compounds in undoped ZnO and Mn doped ZnO.

SEM photograph shown in Fig. 4 shows the size and distribution of particles in the samples. We observe for undoped ZnO the grain size is bigger than the Mn doped ZnO. The SEM investigations of all samples revealed that the crystallites are of nanometer size. The grain growth of ZnO is hindered upon Mn doping, but no significant changes of the grain size is observed from SEM images as the Mn concentration is increased.

For FTIR analysis the KBr pellets are prepared from the undoped and different mol% of Mn doped ZnO powders at two set of temperatures $100^{o}C$ and $500^{o}C$, and their spectrum were shown in Fig. 5(a) and Fig.5(b) respectively. The $100^{o}C$ dried samples retains some typical features of an acetate salt. Two principal absorption peaks are observed between 1650 and 1400 $cm^{-1}$ corresponding to the asymmetric and symmetric stretching of the carboxyl group (C=O). The broad absorption peaks around 3200 $cm^{-1}$ and 2550 $cm^{-1}$ are due to O-H stretching and peaks around 2900 $cm^{-1}$ is due to C-H (acetate) stretching. The absorption peaks observed between 2300-2400 $cm^{-1}$ is because of the existence of $CO_2$ molecule in air. The deformation bands of C=O can also be observed around 1000 $cm^{-1}$ [18]. The $500^{o}C$ annealed samples shows drastic diminishing of carboxylate group (C=O) and the characteristics peak for the C-H (acetate) group, suggesting the loss of acetic acid. Thus on annealing we only observe a very strong band below 500 $cm^{-1}$ due to the Zn-O and (Zn,Mn)-O stretching modes.

The optical transmittance of the undoped ZnO and Mn doped ZnO thin films was determined by the spectrophotometer within the wavelength range of 200-2500nm. For, the transmittance measurements, the films were grown on quartz substrate and irradiated at a perpendicular angle of incidence with quartz glass as reference. The typical room temperature transmittance spectra for undoped ZnO and different concentration of Mn doped ZnO are shown in Fig.6. The optical absorption coefficient ($\alpha$) is evaluated from the transmission spectra using the relation

$$\alpha = -1/d \ln(T) \qquad (1)$$

where d is the thickness of the film and T is the transmittance. The optical band gap was evaluated using the relation:

$$(\alpha h\nu)^2 = A(h\nu - E_g) \qquad (2)$$

where A is a constant, h$\nu$ the photon energy, and $E_g$ is the energy gap. The absorption coefficient was a function of photon energy for undoped and Mn doped ZnO films. The energy gap, $E_g$, is estimated from the intercept of the linear portion of the curve. We observe red shift in the band gap due to Mn doping in ZnO. We obtained the band gap to be 3.27eV for undoped ZnO and it starts decreasing for 1 mol%, 2 mol% and 3mol% of Mn doped ZnO samples as 3.06eV, 2.90eVand 2.78eV respectively. The decrease in $E_g$ for increasing Mn content is attributed to the s-d and p-d interactions giving rise to band gap bowing and it has been theoretically explained using second-order perturbation theory [16].

## 4. Conclusion

ZnO powders containing transition metal Mn synthesized by a simple solvo-thermal process correspond to a hexagonal structure similar to that of undoped ZnO. The XRD measurement suggests that Mn atoms substitute Zn sites in the crystals without changing the wurtzite structure, but with the lattice parameters varying slightly with the extent of doping. We also observed on doping the grain size reduces drastically reducing to nano-scale i.e., doping hinders the grain growth . No secondary phases were observed for the simple synthesis process adapted in the present work for the doped ZnO samples upto 18.31 wt% of Mn doping. The FTIR analysis confirms the formation of ZnO and Mn doped ZnO. The UV-VIS-NIR measurements show the reduction in the band-gap upon Mn doping for concentration of Mn $\leq$ 3 mol%.

**Table 1**: The lattice constants calculated from XRD data of ZnO (sample a)and different mole% of Mn doped ZnO (samples b,c and d)

| Lattice Constants | ZnO | Different mol% Mn doped ZnO | | |
|---|---|---|---|---|
| | | 1 mol% Mn | 2 mol% Mn | 3 mol% Mn |
| a Å | 3.248 | 3.255 | 3.270 | 3.277 |
| c Å | 5.204 | 5.212 | 5.236 | 5.243 |

**Table 2**: Average grain size as a function of Mn in ZnO annealed at $500^{o}$C for 1h.

| Samples | Grain Size (D) nm |
|---|---|
| Undoped ZnO | 19.11 |
| 1 mol% Mn doped ZnO | 15.45 |
| 2 mol% Mn doped ZnO | 17.88 |
| 3 mol% Mn doped ZnO | 18.24 |

**Table 3**:  The wt% are calculated for undoped (sample a) and Mn doped ZnO of 1mol%, 2mol% and 3mol% (samples b,c and d)

| Element | a wt% | b wt% | c wt% | d wt% |
|---------|-------|-------|-------|-------|
| O | 14.93 | 14.87 | 14.85 | 14.34 |
| Mn | --- | 6.54 | 11.59 | 18.31 |
| Zn | 85.07 | 78.58 | 73.56 | 67.35 |
| Total | 100.00 | 100.00 | 100.00 | 100.00 |

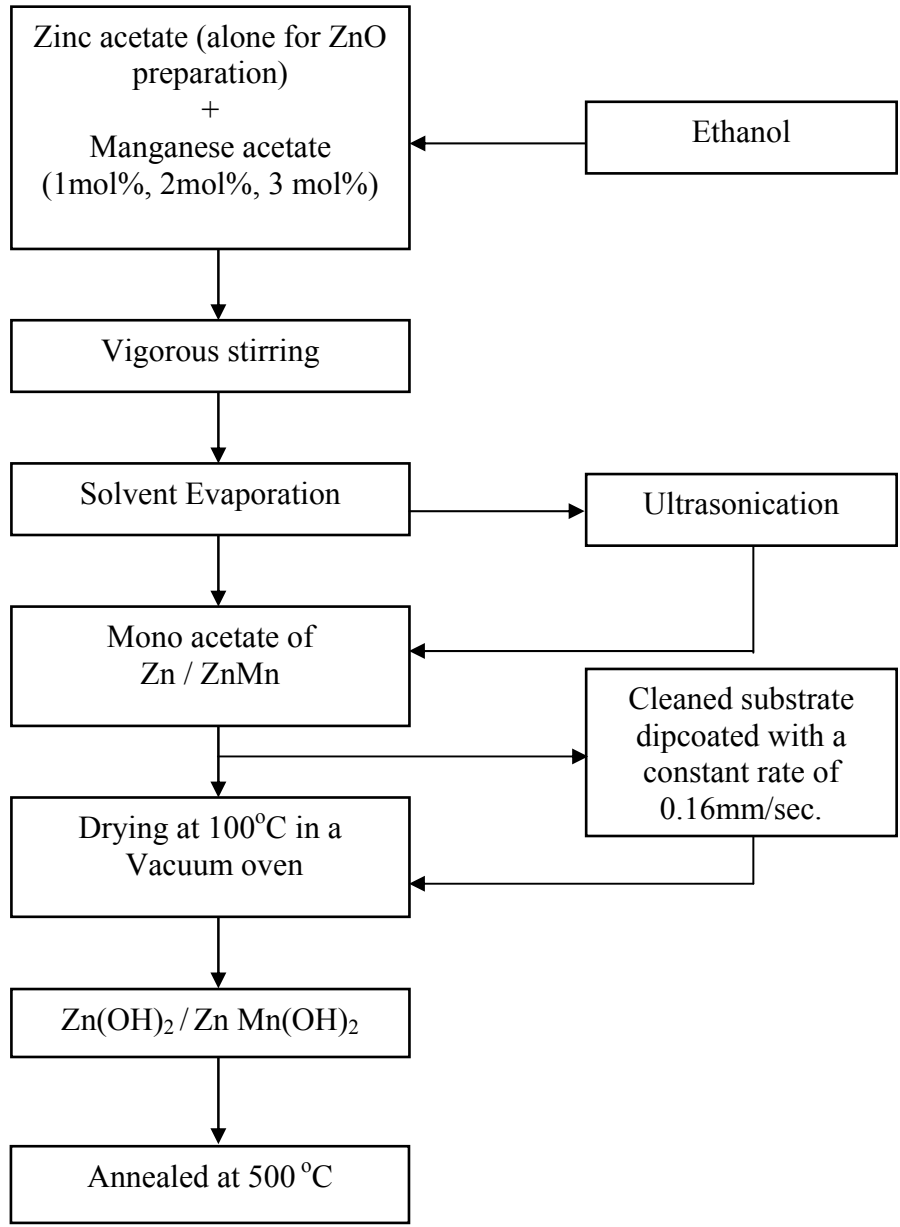

**Figure 1:**  Flow chart describing synthesis of ZnO and Mn doped ZnO

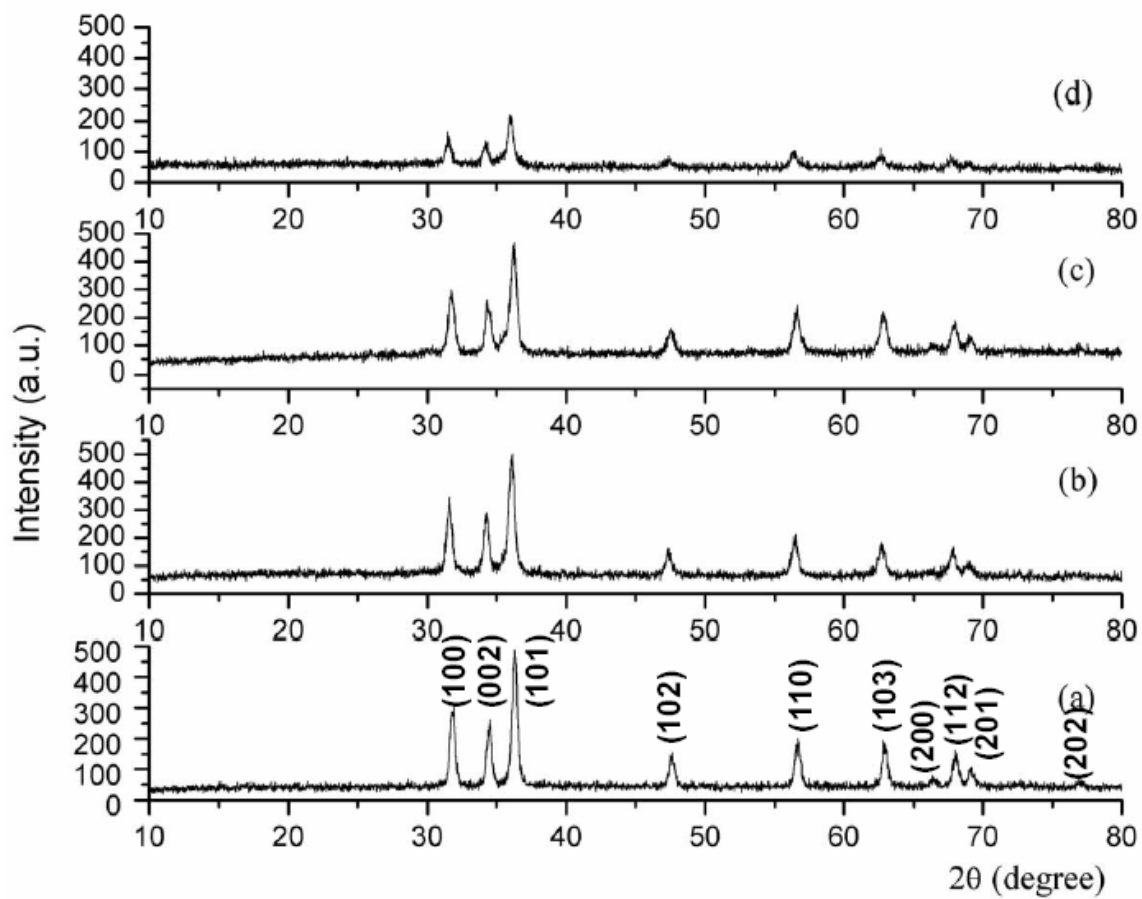

**Figure 2:** XRD patterns of samples annealed in ambient air at 500$^o$C of 1h. (a) undoped ZnO; (b) 1mol% Mn doped ZnO; (c) 2 mol% Mn doped ZnO and (d) 3 mol% Mn doped ZnO.

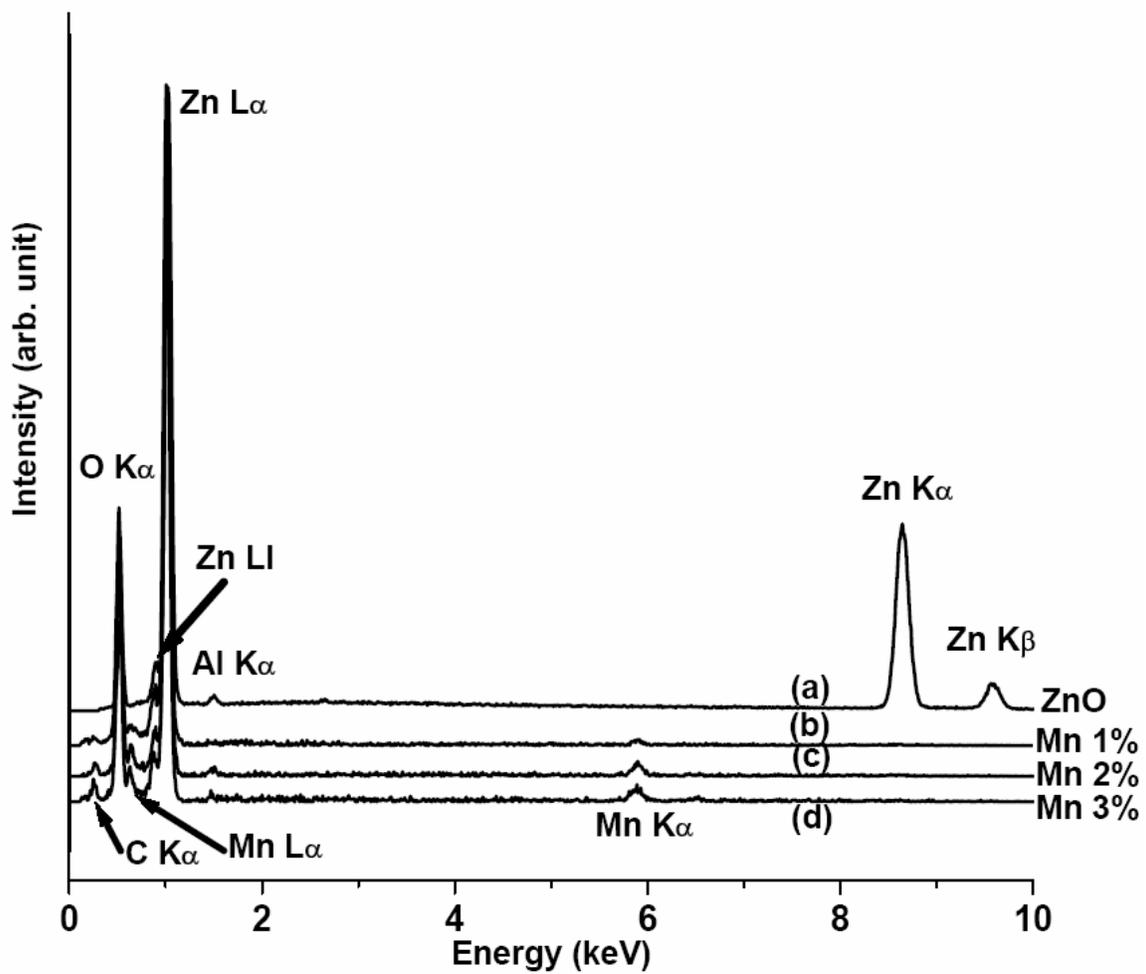

**Figure 3**: EDX of samples (a-d) annealed in ambient air at 500°C of 1h.

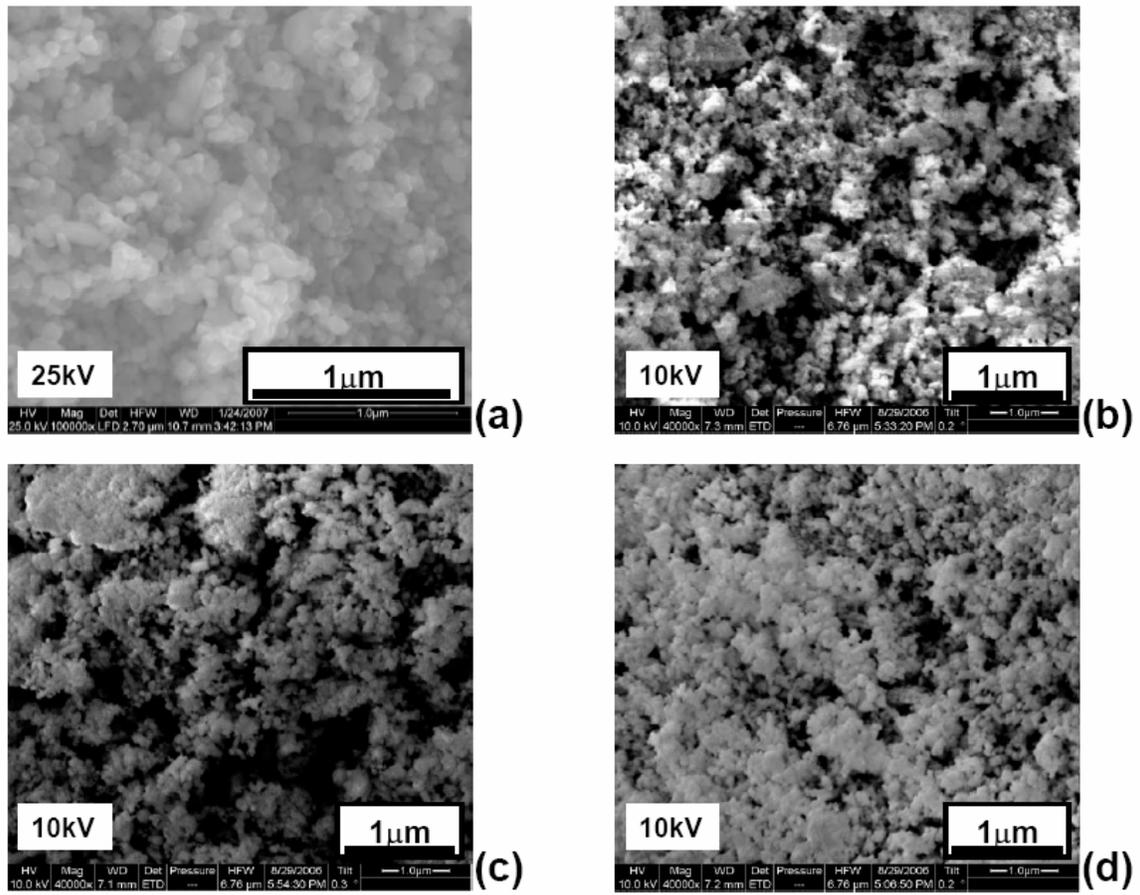

**Figure 4:** Microstructures of undoped and Mn doped ZnO annealed in ambient air at 500ºC for 1 h: (a) undoped ZnO; (b) 1mol% Mn doped ZnO; (c) 2 mol% Mn doped ZnO and (d) 3 mol% Mn doped ZnO.

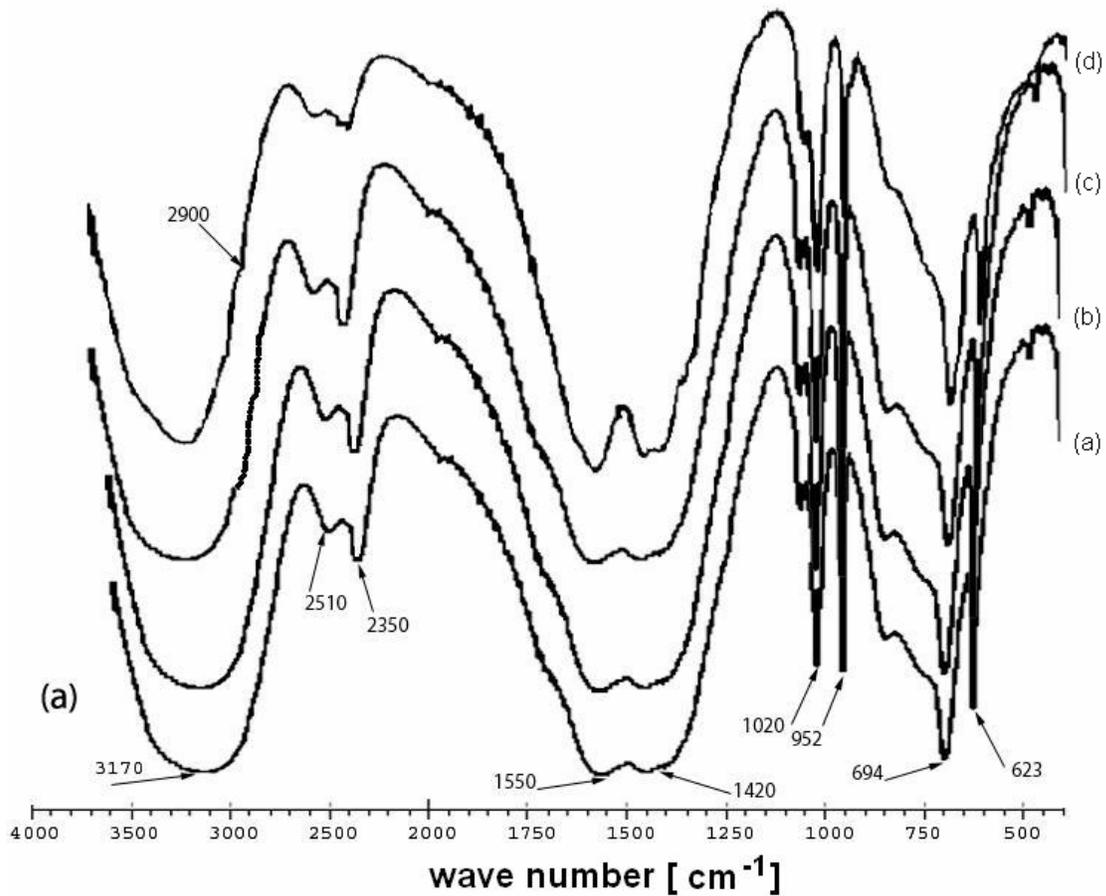

**Figure 5(a):** FTIR spectra of 100°C dried (a) undoped ZnO and Mn doped ZnO: (b)1mol% (c) 2mol% (d) 3mol%

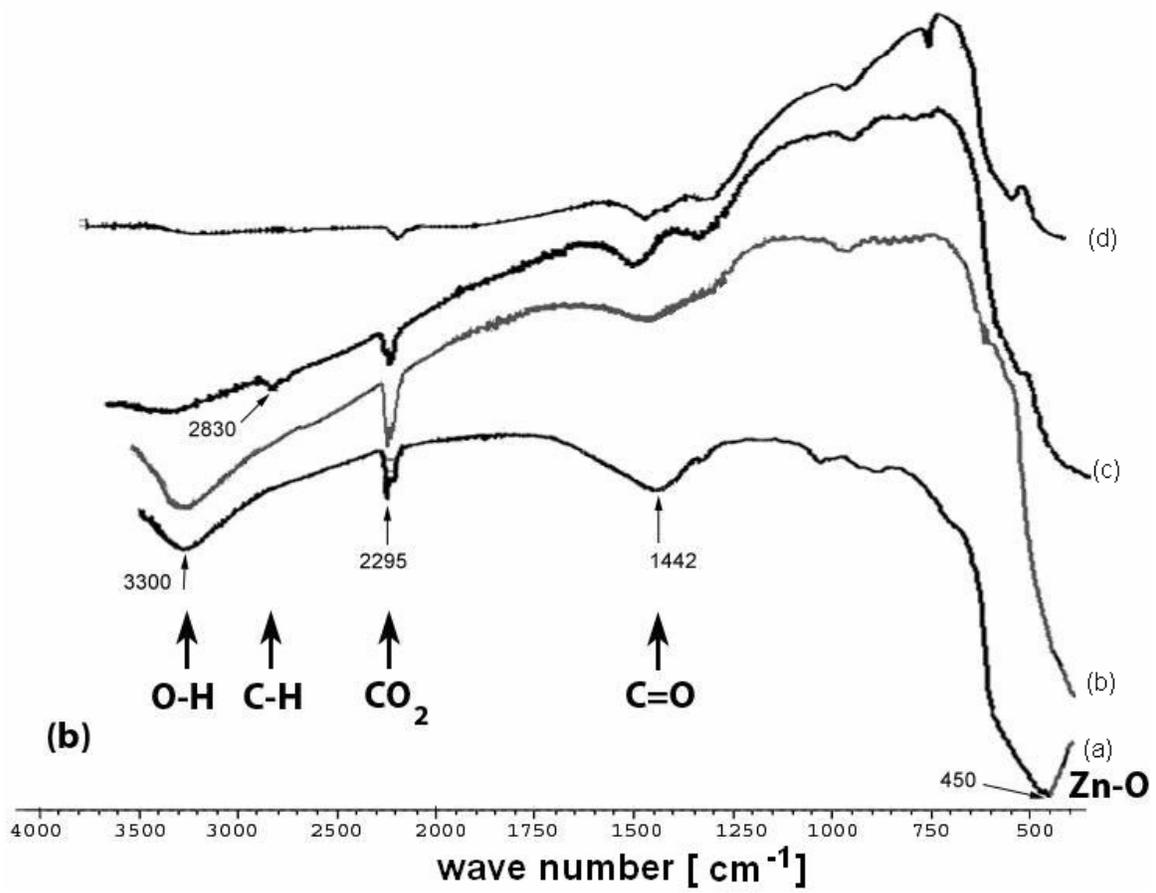

**Figure 5(b):** FTIR spectra of 500°C annealed (a) undoped ZnO and Mn doped ZnO at (b)1mol% (c) 2mol% and (d) 3mol%

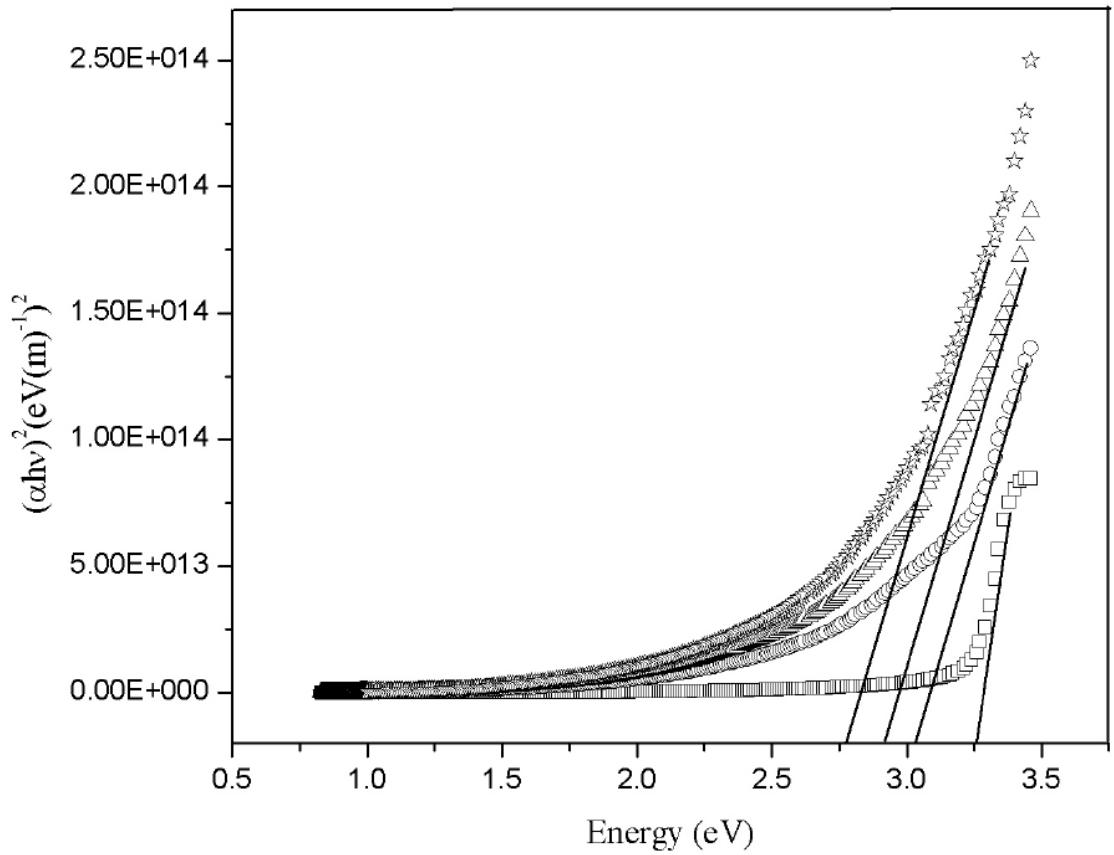

**Figure 6:** Absorption spectra of: (□) undoped ZnO; (○) 1mol% Mn doped ZnO; (△) 2mol% Mn doped ZnO; and (☆) 3 mol% Mn doped ZnO.